\documentclass{PoS}

\usepackage{wrapfig}
\usepackage{graphicx}

\def\figdir{./}

\newcommand\figwidth{.48\textwidth}
\newcommand\Eq[1]{Eq.~\ref{eq:#1}}
\newcommand\Fig[1]{Fig.~\ref{fig:#1}}

\newcommand\calO{\mathcal O}
\newcommand\calC{\mathcal C}
\newcommand\calN{\mathcal N}

\title{Signal/noise optimization strategies for stochastically estimated correlation functions}

\ShortTitle{Signal/noise optimization strategies for stochastically estimated correlation functions}

\author{William Detmold  \\
Center for Theoretical Physics, Massachusetts Institute of Technology, Cambridge, Massachusetts 02139, USA \\
E-mail: \email{wdetmold@mit.edu}}

\author{\speaker{Michael G. Endres} \\
Center for Theoretical Physics, Massachusetts Institute of Technology, Cambridge, Massachusetts 02139, USA \\
E-mail: \email{endres@mit.edu}}

\abstract{
Numerical studies of quantum field theories usually rely upon an accurate determination of stochastically estimated correlation functions in order to extract information about the spectrum of the theory and matrix elements of operators.
The reliable determination of such correlators is often hampered by an exponential degradation of signal/noise at late time separations.
We demonstrate that it is sometimes possible to achieve significant enhancements of signal/noise by appropriately optimizing correlators with respect to the source and sink interpolating operators, and highlight the large range of possibilities that are available for this task.
The ideas are discussed for both a toy model, and single hadron correlators in the context of quantum chromodynamics.
}

\FullConference{The 32nd International Symposium on Lattice Field Theory\\
		 23-28 June, 2014\\
		 Columbia University, New York, NY}

\begin{document}

Monte Carlo simulations of quantum field theories rely heavily upon the reliable stochastic estimation of Euclidean space correlation functions.
Two point correlators, for example, contain information about the spectrum of the theory, whereas three point functions may be used to extract information about matrix elements of operators.
In practice, such correlators often exhibit an exponential degradation of signal to noise at late times, making the extraction of properties of the system challenging.
The canonical example for such difficulties is the determination of the nucleon mass from a two point correlation function.
According to an argument by Lepage \cite{Lepage:1989hd}, the signal/noise of the nucleon correlator decays at the rate $m_N - \frac{3}{2}m_\pi$ at late times, where $m_N$ and $m_\pi$ are the nucleon and pion mass, respectively.

In this work (for full details see \cite{Detmold:2014hla}), we focus on the character of signal/noise for two-point functions of the form
\begin{eqnarray}
C_{ij}(\tau) = \langle \Omega | \hat\calO^\prime_i e^{- \hat H \tau} \hat\calO^\dagger_j| \Omega \rangle
             = \sum_n Z^\prime_{in} Z^*_{jn} e^{-E_n \tau}\ ,
\label{eq:correlator}
\end{eqnarray}
where $\hat H$ is the Hamiltonian of the system, with eigenstates $|n\rangle$ and eigenvalues $E_n$, ordered such that $E_n \le E_{n+1}$, and $| \Omega \rangle$ is the vacuum state.
The labels $i = 1,\cdots,N^\prime$ and $j = 1,\cdots,N$ specify the various sink and source operators, $\hat\calO^\prime_i$ and $\hat\calO_j$, taken to have like quantum numbers.
Note that by inserting a complete set of states, $|n\rangle$, the correlator can in turn be expressed as a sum of exponentials, with overlap factors given by $Z_{in}^\prime = \langle\Omega| \hat\calO^\prime_i |n\rangle $ and $Z_{jn} = \langle\Omega| \hat\calO_j| n\rangle$.

In a numerical simulation, one often constructs a stochastic estimate of the correlator, given by the average, $C = \langle \calC \rangle$, over an ensemble of individual correlators, $\calC$, measured on $\calN$ background field configurations generated by some Markov process.
Let us consider a single correlator formed by the inner product, ${\psi^\prime}^\dagger C\psi$, where $\psi^\prime$ and $\psi$ are complex, unit norm vectors.
These vectors are $N^\prime$ and $N$ dimensional, respectively, and specify a particular linear combination of interpolating operators at the source and sink.
The signal to noise ratio for the estimate, up to a $1/\sqrt{\calN}$ proportionality constant, is given by 
\begin{eqnarray}
\theta_c({\psi^\prime},\psi) = \left[ \frac{1}{ \theta^2({\psi^\prime},\psi)} -1  \right]^{-1/2} \ ,\qquad \theta({\psi^\prime},\psi) = \frac{\left| {\psi^\prime}^\dagger C \psi\right| }{ \sigma({\psi^\prime},\psi) }\ ,
\label{eq:signal_noise}
\end{eqnarray}
where
\begin{eqnarray}
\sigma^2({\psi^\prime},\psi) = \left({\psi^\prime} \otimes{\psi^\prime}^* \right)^\dagger \Sigma^2 \left( \psi \otimes\psi^* \right)\ ,\qquad  \Sigma^2 = \langle \calC \otimes \calC^* \rangle \ .
\label{eq:Sigma}
\end{eqnarray}
Since in most cases of interest $\theta_c \approx \theta$, we refer to both $\theta_c$ and $\theta$ as the ```signal/noise''.

\begin{wrapfigure}{R}{0.5\textwidth} 
\centering
\includegraphics[width=\figwidth]{\figdir 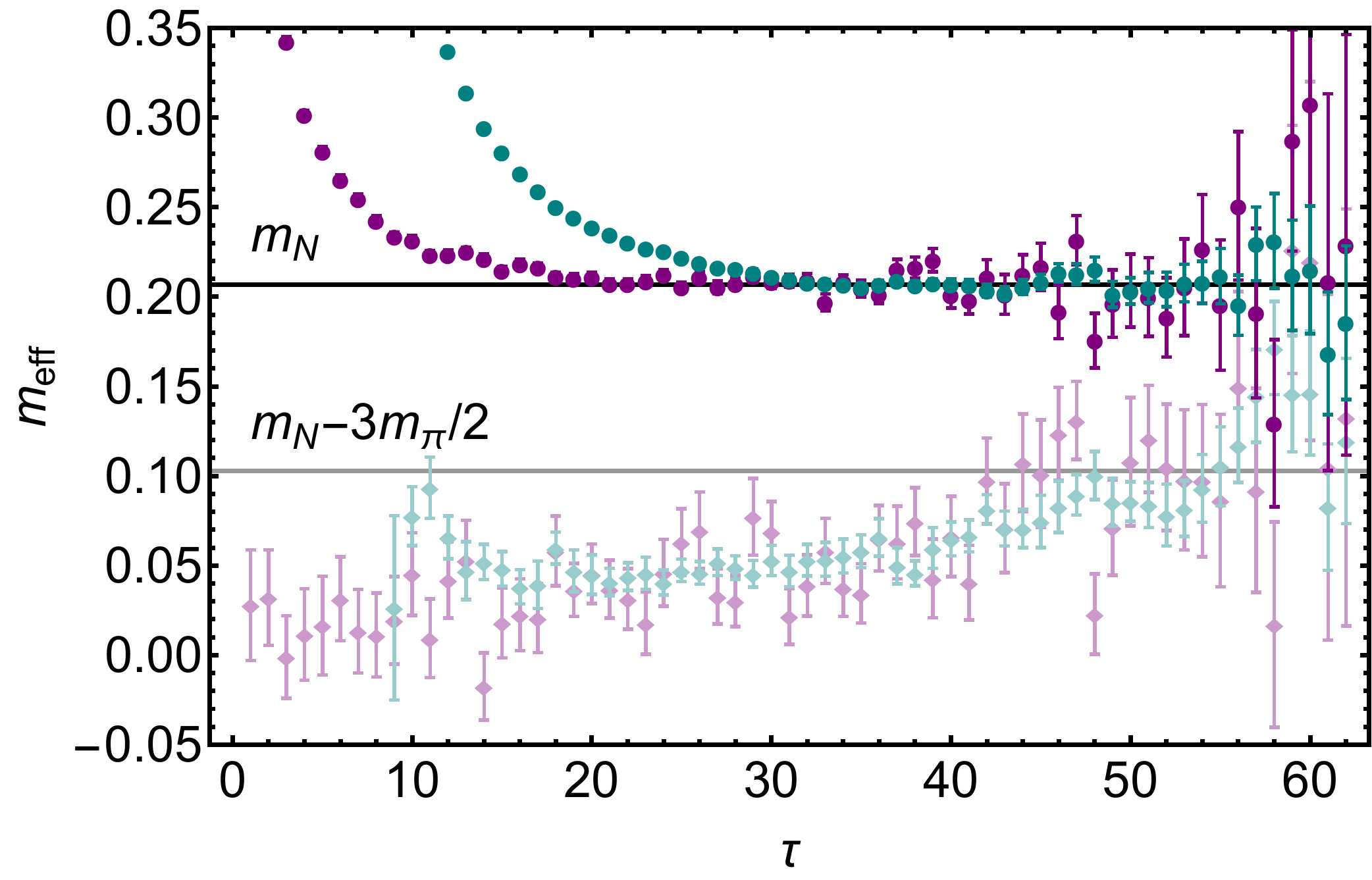}
\caption{\label{fig:nucleon_meff}%
Effective mass plots for the nucleon correlator (circles) and effective mass plots for the associated signal noise (diamonds), for two different source and sink interpolating fields.
}
\end{wrapfigure}

By studying the late-time exponential decay of the correlation functions, one is in principle able to extract information about the low-lying spectrum of the theory.
In particular, it is common practice to consider the late-time behavior of the  effective mass, defined as
\begin{eqnarray}
m_{eff}(\tau) = -\frac{1}{\Delta\tau} \log \frac{{\psi^\prime}^\dagger C(\tau+\Delta\tau) \psi}{ {\psi^\prime}^\dagger C(\tau) \psi}
\to E_0 + \left( \frac{{\psi^\prime}^\dagger Z_1^\prime Z_1^\dagger \psi}{{\psi^\prime}^\dagger Z_0^\prime Z_0^\dagger \psi } \right) \frac{1-e^{-\Delta E_1\Delta\tau}}{\Delta\tau} e^{-\Delta E_1 \tau} + \cdots\ ,
\label{eq:effective_mass}
\end{eqnarray}
for $\Delta E_1 = E_1 - E_0$ and some $\Delta\tau$, often taken to be a single lattice spacing.
For general choices of $\psi^\prime$ and $\psi$, the effective mass yields a constant, the ground state energy, up to exponentially suppressed excited state contamination.
By tuning the source and sink vectors so that they are orthogonal to $Z_1$, and/or $Z_1^\prime$, one can reduce such contamination.
\Fig{nucleon_meff} shows examples of an effective mass, plotted for the nucleon two-point correlator, as a function of the time separation between source and sink interpolating fields.
Examples are shown for a Hermitian matrix of correlators with two choices of $\psi^\prime=\psi$: one that is optimized to produce an early plateau, and one that is not.
In each case, excited state contamination dominates at early times, a plateau corresponding to the ground state emerges at intermediate times, and noise dominates at late times.
In the same figure, we plot corresponding effective masses for the signal/noise and find that they indeed tend toward the expected value, $m_N-\frac{3}{2} m_\pi$, at late times.
Because of the finite temporal extent of the lattice, the signal/noise degradation can in fact be even worse than this expectation \cite{Beane:2009gs}.

In order to maximize the plateau region over which energies may be extracted (via a least-squares fit), one has two options: either reduce the signal/noise, thereby extending the plateau at late times, or alternatively, find linear combinations of interpolating operators which maximize the overlap onto energy eigenstates, thereby extending the plateau to earlier times.
Increasing the ensemble size can achieve the former, but only with logarithmic improvement.
Well established methods exist for the latter; for Hermitian correlators, these go by the names ``variational method'' or ``generalized eigenvalue problem'' \cite{Michael:1982gb,Michael:1985ne,Luscher:1990ck,Blossier:2009kd}; in the case of nonsymmetric correlators or correlators formed from a limited basis, the ``matrix-Prony'' or ``generalized pencil-of-function'' methods may be used \cite{Beane:2009gs,Beane:2009kya,Fleming:2004hs,Aubin:2011zz}.
In the nucleon example shown here, one finds that performing such an optimization indeed yields a plateau at earlier times, but at the cost of {\it enhanced} statistical uncertainties compared to the typical un-optimized source.
The result illustrates how signal/noise is not only influenced by the ensemble size, but also the choice of interpolating fields.
Furthermore, it suggests an inherent incompatibility between reducing excited state contamination at early times and reducing statistical uncertainties on the correlator at late times.
Since uncertainties on extracted energies depend on numerous factors, including the type of fit, fit interval and uncertainties on the correlator itself, it is {\it a priori} unclear which strategy is ideal.

\begin{wrapfigure}{R}{0.5\textwidth} 
\centering
\includegraphics[width=\figwidth]{\figdir 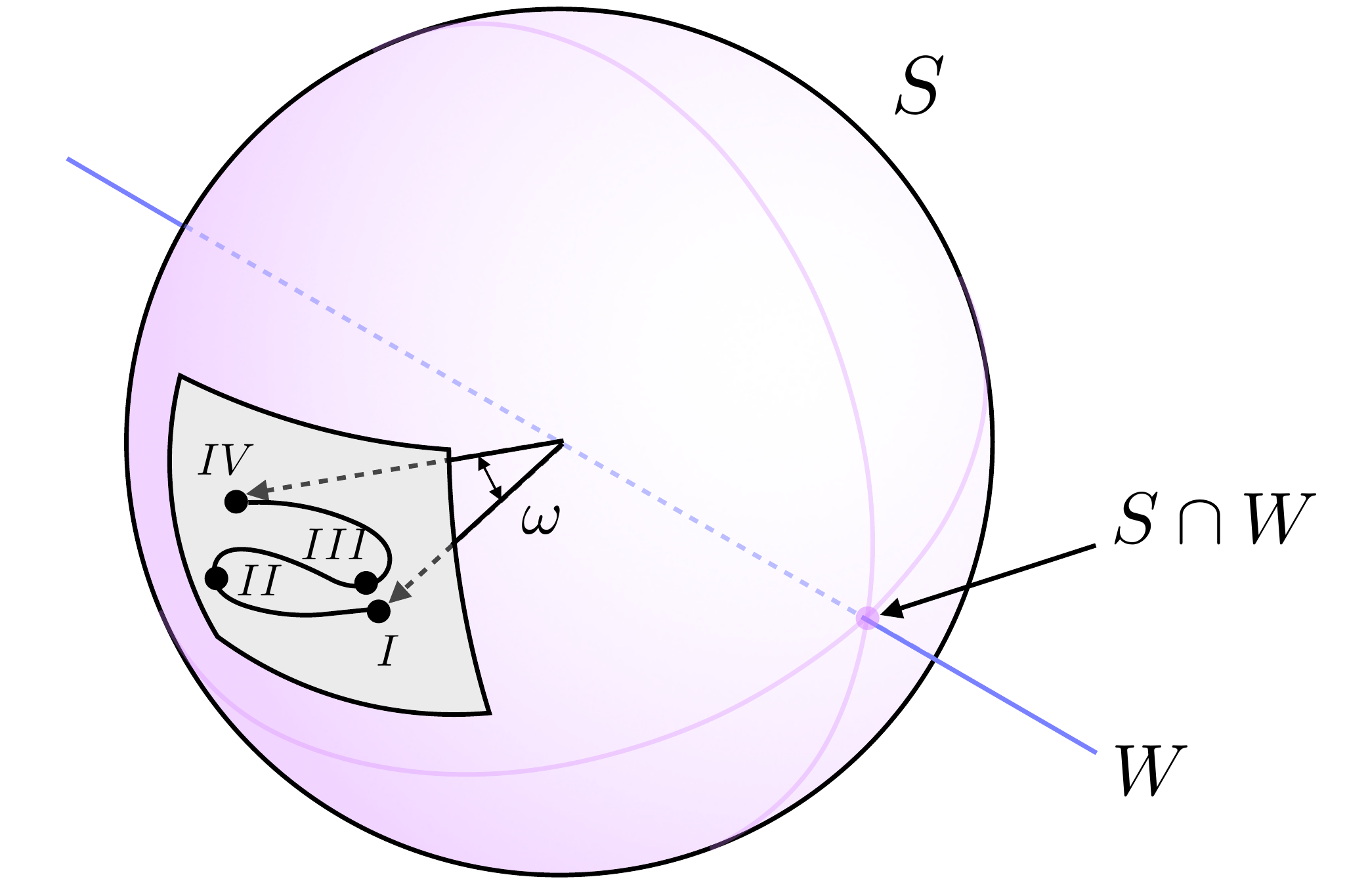}
\caption{\label{fig:landscape1}%
Signal/noise landscape as a function of $\psi^\prime$, for a fixed $\psi$.
Although not shown, there exist one ``flat direction'' on this landscape, corresponding to phase rotations of $\psi^\prime$.
}
\end{wrapfigure}

\begin{figure}[b] 
\centering
\includegraphics[width=\figwidth]{\figdir 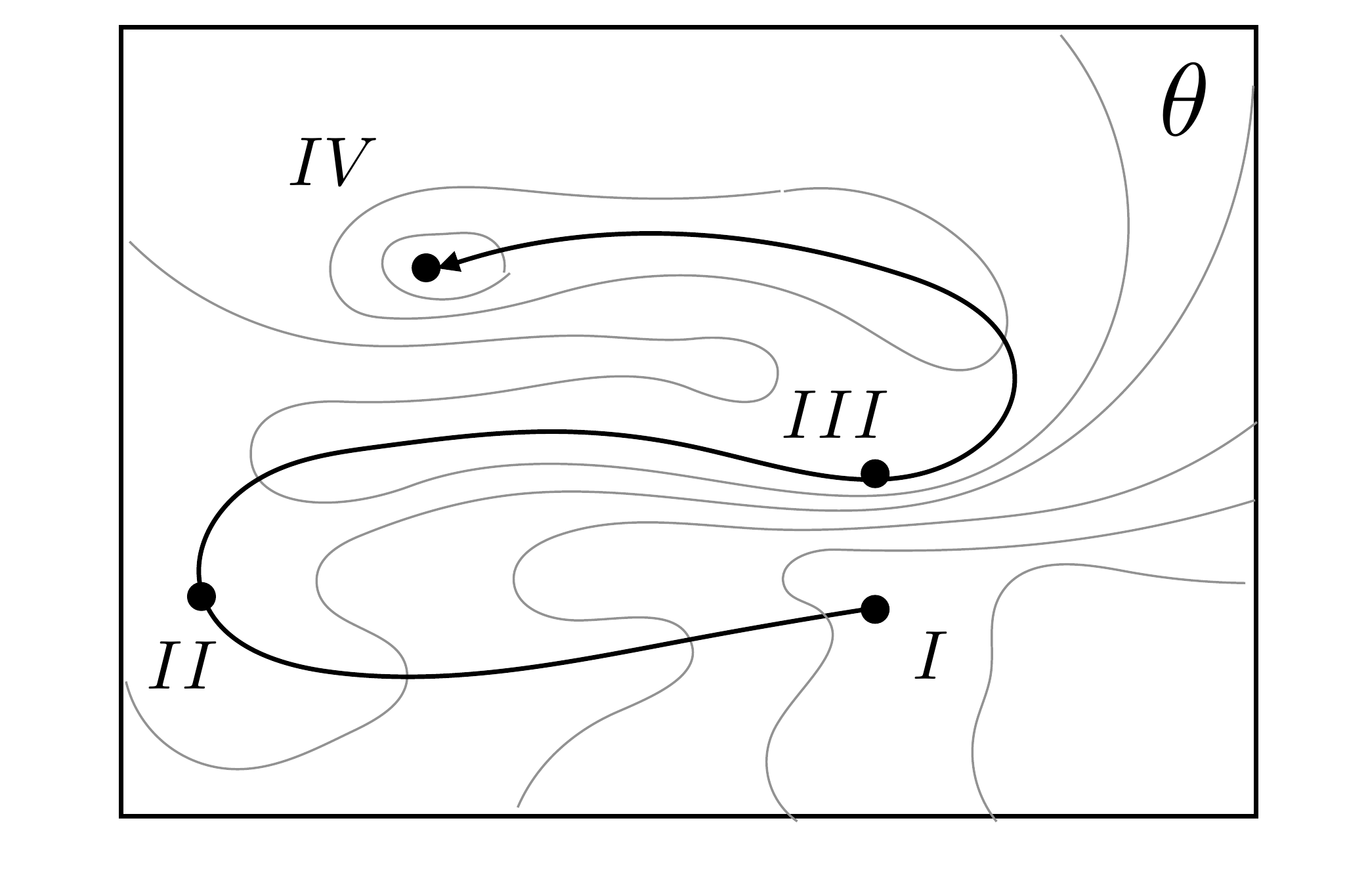}
\includegraphics[width=\figwidth]{\figdir 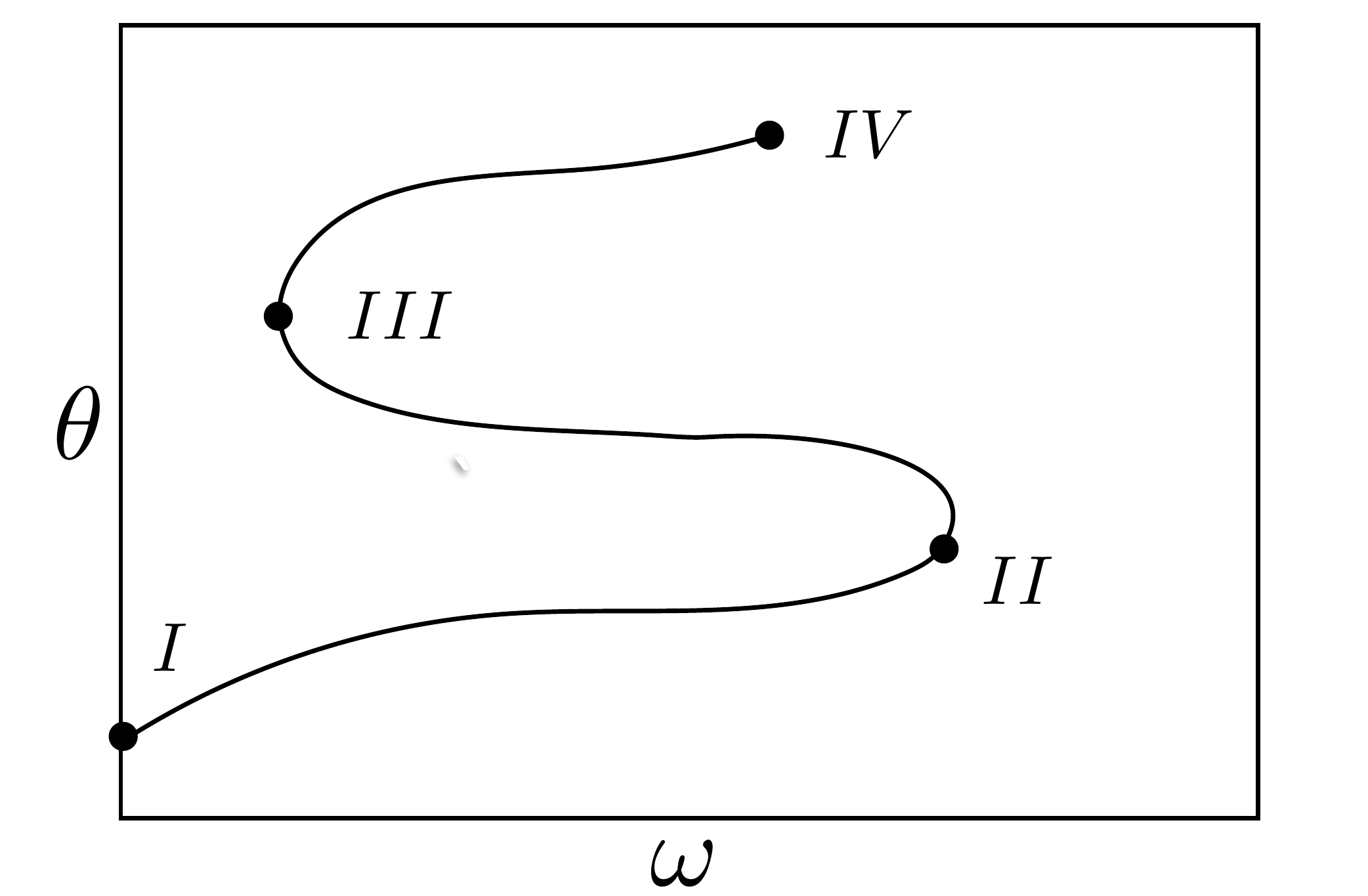}
\caption{\label{fig:landscape2}%
Left: Signal/noise landscape on a patch of $S$, displayed in Fig.~2.
Right: Signal/noise as a function of the overlap angle $\omega$ along a path of steepest ascent from point (I) to point (IV).
}
\end{figure}

To gain further insight, it is instructive to imagine how the signal/noise at late times behaves as a function of the interpolating fields.
Consider, as one example, the signal/noise ``landscape'' as a function of the sink vector, $\psi^\prime$, for some fixed source vector $\psi$.
For the moment, let us forget that $\psi^\prime$ has unit norm, and allow the signal/noise to be defined on $\mathbb{R}^{2N^\prime}$.
There exists a $2N^\prime-2$ dimensional subspace, $W$, for which $\theta(\psi^\prime,\psi)$ exactly vanishes, and it is given by the set of $\psi^\prime$ orthogonal to $C\psi$.
Note that this subspace is continuously connected to the origin, $\psi^\prime=0$.
The unit norm constraint confines $\psi^\prime$ to a $(2N^\prime-1)$-sphere ($S$) embedded within $\mathbb{R}^{2N^\prime}$, and which intersects $W$.
A schematic of this geometry is provided in \Fig{landscape1}.

In \Fig{landscape2} (left), we show a patch of the signal/noise landscape defined on $S$.
The eigenstate-optimized sink vector can lie anywhere on the domain, such as at point $(I)$ in the figure.
Since $S$ is compact, one expects a global maximum of the signal/noise, such as point $(IV)$, at some other location on the domain.
One can show that these two points are continuously connected along a path of steepest ascent, passing through the intermediate points (II) and (III).
Generally speaking, the overlap between the vectors (I) and (IV), characterized by the overlap angle $\omega$ (indicated in \Fig{landscape1}), can be either large or small, and is completely independent of the change in the signal/noise between the two points.
The implications of this observation are rather intriguing: a correlator obtained by using the eigenstate-optimized sink vector (I) can have {\it arbitrarily} poor signal/noise compared to that of (IV).
Furthermore, point (I) may lie arbitrarily close to $S\cap W$, where the signal/noise vanishes.
In perhaps the most severe of unfortunate scenarios (not shown in the figure), point (I) may even lie at the bottom of a precipice, whereas point (IV) may lie at the top, only a short distance away as measured by the overlap angle $\omega$.
In that scenario, an enormous enhancement in signal/noise could be possible, going from (I) to (IV), while introducing only a tiny amount of additional excited state contamination to the correlator.
The range of possibilities is vast, and highly dependent on both the system under study and basis of interpolating operators involved. 

Given the potentially nontrivial nature of the landscape, it is important to consider the interplay between excited state contamination and signal/noise in correlators, particularly as one travels along the path of steepest ascent.
A schematic example of signal/noise, as a function of $\omega$, is provided in \Fig{landscape2} (right) and corresponds to the route of steepest ascent shown in \Fig{landscape2} (left).
Although point (IV) yields the greatest signal/noise enhancement for the correlator in this example, smaller uncertainties in the extracted energies may be possible at point (III), corresponding to a significantly smaller $\omega$, yet only a moderately diminished $\theta$.

\begin{figure} 
\centering
\includegraphics[width=\figwidth]{\figdir 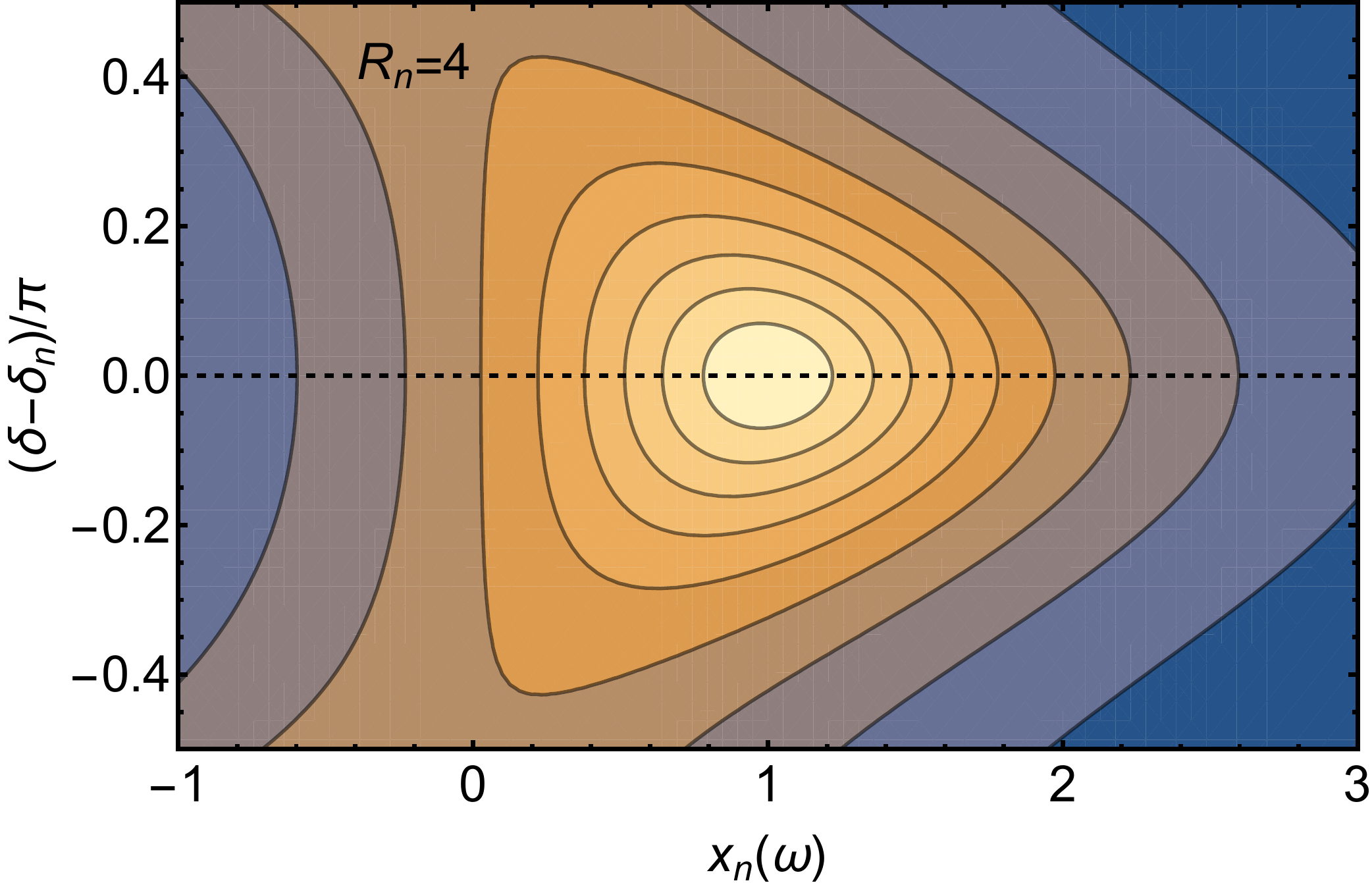}
\includegraphics[width=\figwidth]{\figdir 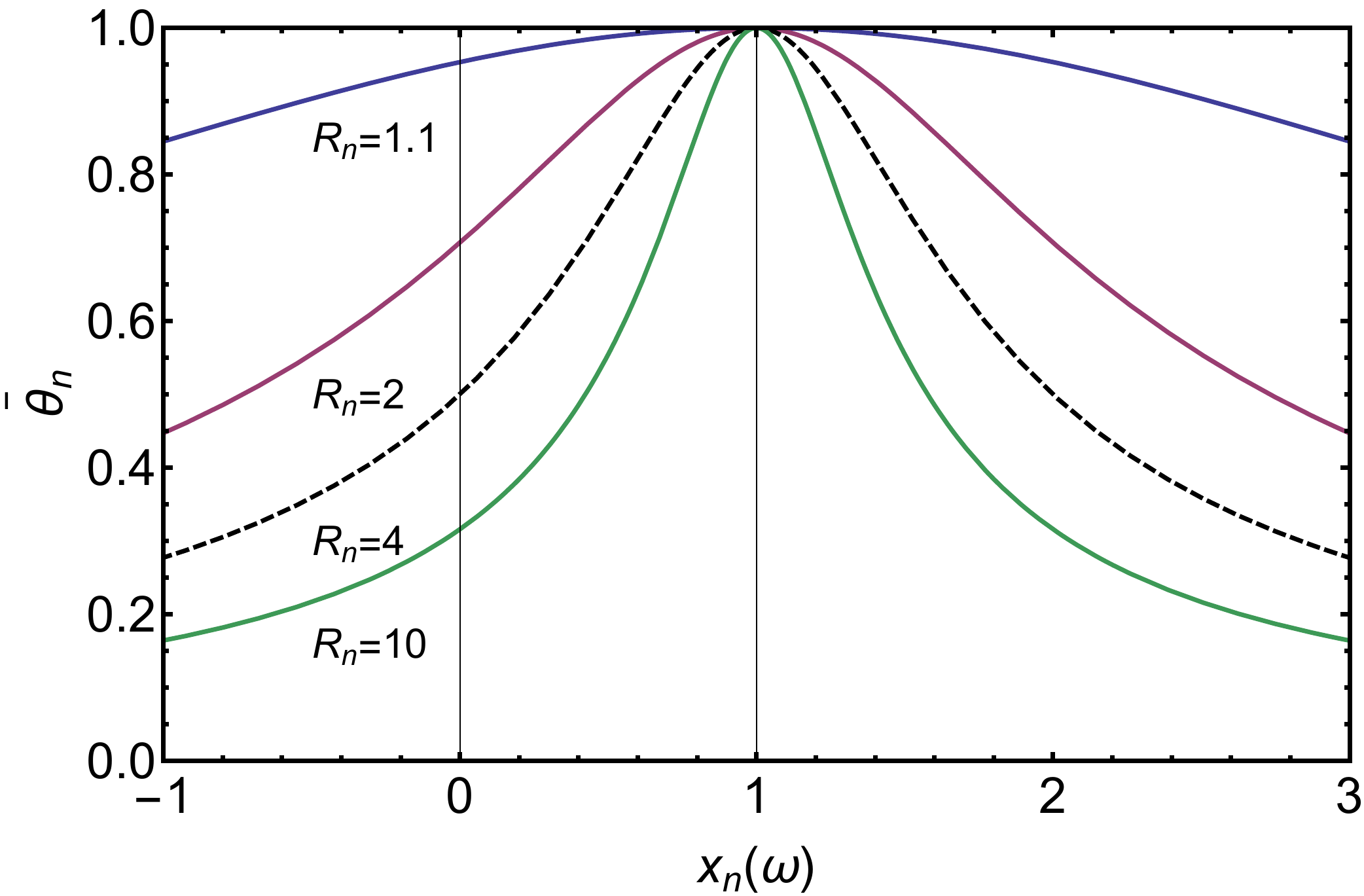}
\caption{\label{fig:toy_ratio}%
Left: Density plot of the normalized signal/noise ratio $\bar\theta_n(\omega,\delta)$ as a function of the sink parameterization angles $(\omega,\delta)$, for $R_n=4$.
Right: Signal/noise as a function of $x_n(\omega)$, for various values of $R_n$.
Dashed curve in each plot corresponds to the same values of $R_n$ and $\delta_n$.
}
\end{figure}

Although the arguments presented thus far have been heuristic, it is possible to make the analysis explicit.
Let us begin by considering an idealized scenario of an arbitrary two-state system.
We consider a correlation function in the eigenbasis of the Hamiltonian, and a source $\psi_n$ ($n=0,1$), which has perfect overlap onto the eigenstate $|n\rangle$.
The correlator ${\psi^\prime}^\dagger C(\tau) \psi_n$ is therefore a pure exponential, with a decay rate $E_n$.
We may fully parameterize the sink vector, up to an overall irrelevant phase factor, by $\psi^\prime(\omega,\delta) = ( \cos(\omega),\sin(\omega) e^{i\delta} )$, where $\omega\in[0,\pi)$, and $\delta\in[-\pi/2,\pi/2)$.
Given a fixed, $\psi_n$, let us define $(\omega_n,\delta_n)$ to be the critical angles at which the signal/noise ratio is maximized.
Then as a function of $(\omega,\delta)$, one can prove on general grounds that the normalized signal/noise ratio, $\bar\theta_n(\omega,\delta)\equiv\theta(\psi^\prime(\omega,\delta),\psi_n)/\theta(\psi^\prime(\omega_n,\delta_n),\psi_n)$, has the functional form
\begin{eqnarray}
\bar\theta_n(\omega,\delta) = \frac{1}{\sqrt{R_n + (R_n-1) x_n(\omega) \left[ x_n(\omega) -2 \cos (\delta - \delta_n) \right] }} \ ,
\label{eq:toy_sn}
\end{eqnarray}
where $\theta(\psi^\prime(\omega_n,\delta_n),\psi_n) \equiv \sqrt{R_n} \theta(\psi_n,\psi_n)$, and
\begin{eqnarray}
x_0(\omega) = \frac{\tan\omega}{\tan\omega_0}\ ,\qquad
x_1(\omega) = \frac{\tan\omega}{\tan\omega_1}\ .
\end{eqnarray}
We may interpret the parameter $\sqrt{R_n}\ge1$ as the amount of enhancement in the signal/noise for a correlator constructed using the sink $\psi^\prime(\omega_n,\delta_n)$, compared to one constructed using the sink $\psi_n$.
The parameter $x_n(\omega)$ is zero when the sink vector equals the source vector, and unity when the signal/noise is maximum.
Interestingly, the functional form of \Eq{toy_sn} is the square-root of the Breit-Wigner formula centered about $x_n(\omega)=1$, and with a half-width at half maximum given by $(R_n-1)^{-1/2}$.
The behavior of \Eq{toy_sn} is shown in \Fig{toy_ratio} for various choices of $R_n$.

As previously noted, $R_n$ and $\omega_n$ are independent parameters, and as such, it is possible for a system to possess a very large $R_n$ and a very small $\omega_n$, thus realizing the ``precipice scenario'' previously discussed.
The toy model analysis presented here can be extended to the case where source and sink vectors are equal and parameterized by $(\omega,\delta)$.
In that case, an additional time-dependent parameter enters into the parameterization of the signal/noise, which provides a measure of the amount of excited state contamination present at the global maximum.
In the late time limit, this parameter vanished exponentially, corresponding to ground state domination, and the functional form of the normalized signal/noise tends to the square of \Eq{toy_sn}.
At intermediate times, a variety of interesting signal/noise enhancement scenarios are also possible \cite{Detmold:2014hla}.

Let us finally turn to a realistic application of the ideas discussed here, focusing in particular on single hadron correlation functions in quantum chromodynamics (QCD).
Here, we determined the grounds state energies of the pion, proton and delta baryon, and the ground and first excited state energies of the rho meson.
The pion, proton and delta baryon energies were extracted from approximately Hermitian $5\times 5$ correlator matrices, closely related to those of \cite{Beane:2009kya}, whereas the rho meson energies were extracted from $26\times 26$ matrices used in \cite{PhysRevD.87.034505}.
All correlators were measured on anisotropic gauge field configurations generated by the Hadron Spectrum Collaboration using a $2+1$ flavor tadpole-improved clover fermion action and a Symanzik-improved gauge action \cite{PhysRevD.78.054501,PhysRevD.79.034502}.
The rho meson correlators were measured on $24^3\times128$ lattices, and the remainder were measured on $20^3\times128$ lattices.
All lattices were generated with an anisotropy $b_s/b_\tau \approx 3.5$, where $b_s = 0.1227(8)$ fm and $b_\tau$ are the spatial and temporal lattice spacings (hereafter set to unity).
Quark masses for these ensembles correspond to a pion mass, $m_\pi \approx 390$ MeV, and kaon mass, $m_K\approx 546$ MeV. 
Pion, proton and delta baryon correlators were computed on an $\calN=305$ ensemble using $\calO(30)$ randomly placed Gaussian-smeared sources and zero-momentum projected Gaussian-smeared sinks, creating a stochastically approximated wall source.
Rho meson correlators were computed on an $\calN=566$ ensemble, and constructed from zero-momentum projected operators belonging to the irreducible representation, $T_1^-$, of the octahedral group with parity.

\begin{figure}
\centering
\includegraphics[width=\figwidth]{\figdir 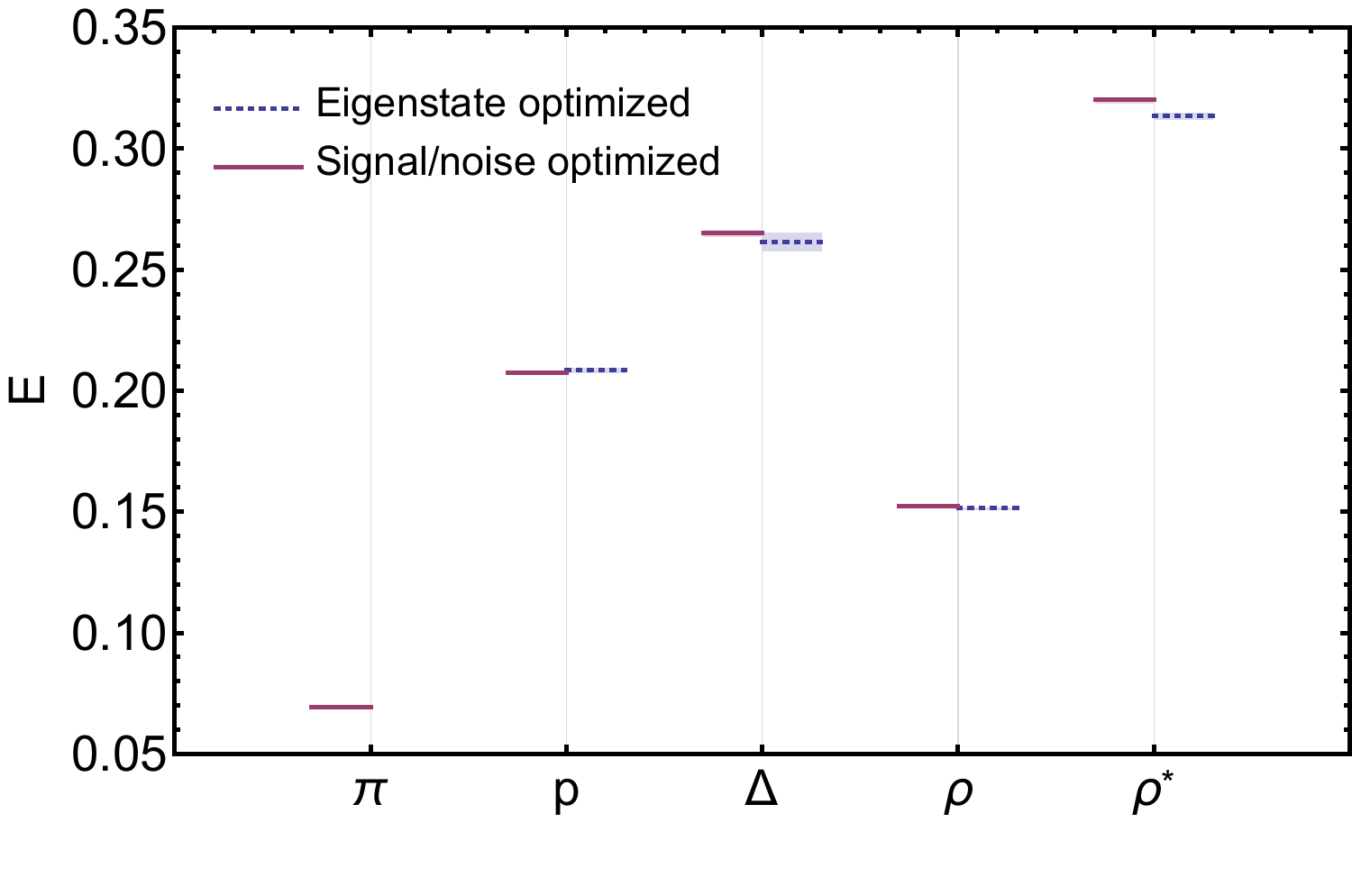}
\includegraphics[width=\figwidth]{\figdir 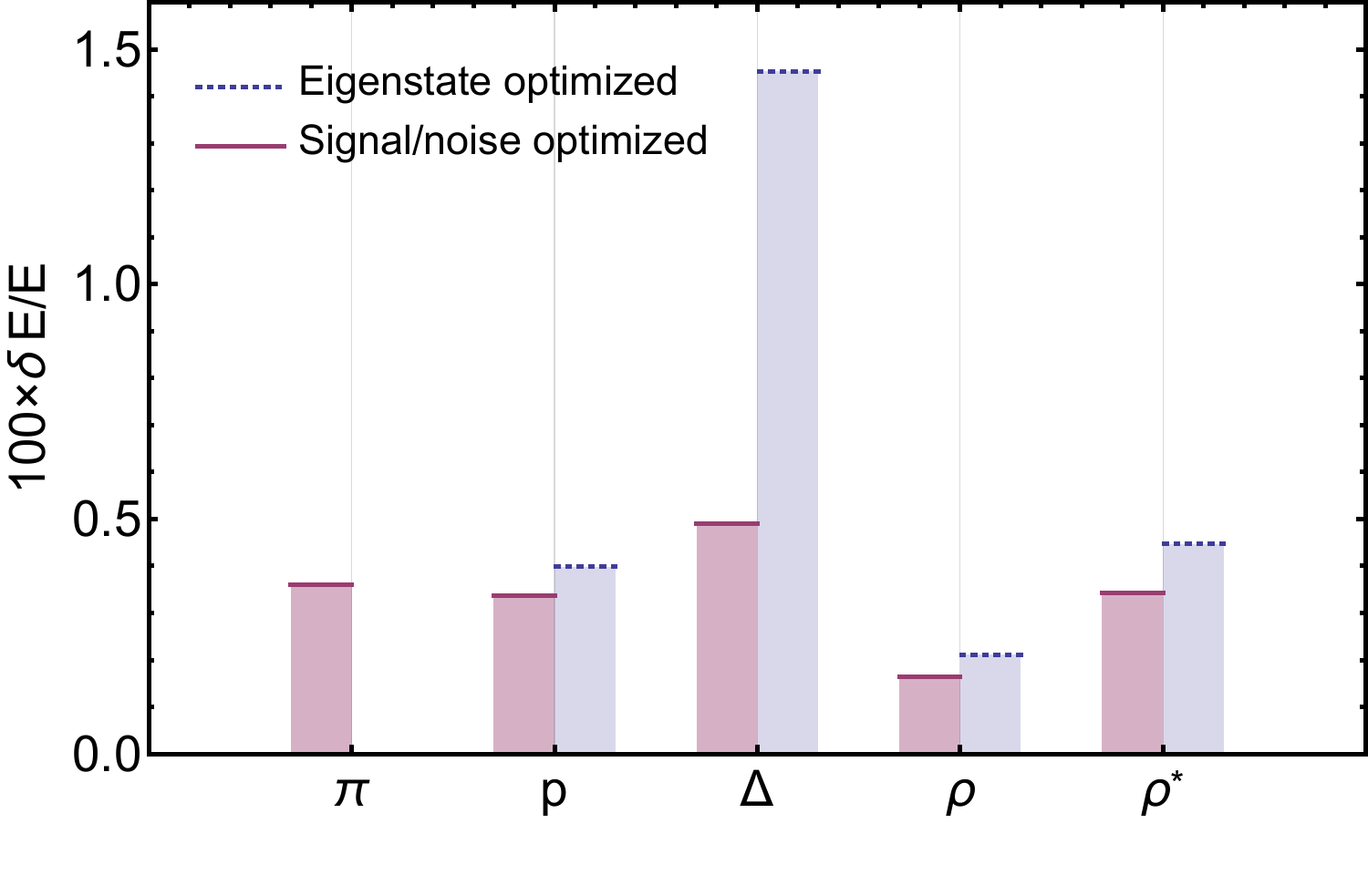}
\caption{\label{fig:summary}%
Left: Extracted energies for the pion ($\pi$), proton ($p$), delta baryon ($\Delta$), rho meson ground ($\rho$) and first excited ($\rho^\star$) states. Energies were obtained from eigenstate and signal/noise optimized correlators.
Right: A comparison of the corresponding relative errors in the extracted energies.
}
\end{figure}

Multi-exponential least-squares fits were performed for the correlators at various locations along a path of steepest ascent on two signal/noise landscapes.
One landscape was defined holding the source fixed and allowing the sink to vary, whereas the second was defined by holding the source and sink equal, and allowing them to vary simultaneously.
In each case, the trajectory started at a point where overlap with an eigenstate was maximum (eigenstate optimized), and ended at a point where the signal/noise was maximum (signal/noise optimized).
Fits were then performed over multiple temporal intervals while holding the upper limit of the interval fixed; fits satisfying a $\chi^2/d.o.f \le 1.1$ were deemed acceptable.
Among the acceptable fits, ones corresponding to the largest fit interval were selected and compared for various points along the trajectory and for each fit model (i.e., one-, two-, and three-exponentials).
\Fig{summary} provides a comparison of the best fit results among all eigenstate optimized and signal/noise optimized correlators (left), and their associated relative errors (right).
With exception to the excited rho state, which possessed significant systematic uncertainties associated with the fit interval, all extracted energies were statistically consistent.
In most cases, only a modest reduction of uncertainties for extracted energies is evident for the signal/noise optimized correlators.
However, in the case of the delta baryon, a three-fold reduction was achieved.

The variety of outcomes achieved for single hadrons illustrates an inherent dependence of the proposed methods on both the properties of the system and choice of operator basis.
It would be particularly interesting to explore whether expanding the basis of operators (particularly at the sink, where the task is computationally inexpensive) might improve the outcome of these results.
It may also be profitable to include operators with different quantum numbers in correlator matrices; although such operators would contribute nothing to the signal, they may lead to nontrivial cancellations in the noise (i.e., \Eq{Sigma}).
Finally, the ideas presented here are quite general, and may prove useful for analyzing multi-nucleon correlators, three-point functions and disconnected diagrams.

\begin{acknowledgments}
We would like to gratefully acknowledge S. Meinel for the use of his QMBF and XMBF fitting software.
We would also like to acknowledge J. Dudek for generously sharing rho meson correlator data from the Hadron Spectrum Collaboration, and K. Orginos and A. Walker-Loud for generously sharing pion, proton and delta baryon correlator data.
The latter data were generated with Teragrid resources and local resources at the College of William and Mary.
This study was supported by the U. S. Department of Energy under cooperative research agreement Contract No. DE-SC0011090, the U. S. Department of Energy Early Career Research Award No. DE-SC0010495, and the Solomon Buchsbaum Fund at MIT.
\end{acknowledgments}

\bibliography{signoise}
\bibliographystyle{h-physrev.bst}

\end{document}